\documentstyle[12pt,epsf]{article}
\begin{document}
\begin{rightline}
{CU-TP-1035}
\end{rightline}
\vskip 20pt

\noindent{\bf PARTON SATURATION-AN OVERVIEW\footnote{Lectures given at the Cargese Summer School, August 6-18, 2001}}\\

\vskip 20pt
\noindent{A.H. Mueller\footnote{This work is supported in part by the Department of Energy}\\
Department of Physics, Columbia University\\
New York, New York 10027}

\vskip10pt
The idea of partons and the utility of using light-cone gauge in QCD are
introduced. Saturation of quark and gluon distributions are discussed
using simple models and in  a more general context.  The Golec-Biernat
W\"usthoff model and some simple phenomenology are described.  A simple,
but realistic, equation for unitarity, the Kovchegov equation, is
discussed, and an elementary derivation of the JIMWLK equation is given.

\section{Introduction}

These lectures are meant to be an introduction, and an overview, of
parton saturation in QCD.  Parton saturation is the idea that the
occupation numbers of small-x quarks and gluons cannot become arbitrarily
large in the light-cone wavefunction of a hadron or nucleus.  Parton
saturation is an idea which is becoming well established theoretically
and has important applications in small-x physics in high-energy
lepton-hadron collisions and in the early stages of high-energy heavy ion
collisions.  The current experimental situation is unclear.  Although
saturation based models have had considerable phenomenological success in
explaining data at HERA and at RHIC more complete and decisive tests are
necessary before it can be concluded that parton saturation has been
seen.  These lectures begin very simply by describing, through an
example, some important features of light-cone perturbation theory in QCD,
and they end by describing some rather sophisticated equations which
govern light-cone wavefunctions when parton densities are very large.

\section{States in QCD Perturbation Theory}

In light-cone perturbation theory states of QCD are described in terms of
the numbers and distributions in momentum of quarks and gluons.  The
essential features of light-cone perturbation theory that will be needed
in these lectures can be illustrated by considering the wavefunction of
a quark through lowest order in the QCD coupling g.  One can write

\begin{equation}
\vert\psi_p> = N\vert p> + \sum_{\lambda=\pm}\sum_{c=1}^{N_c^2-1}\int
d^3k\psi_\lambda^c(k)\vert p-k;k(\lambda,c)>
\end{equation}

\noindent where $\vert p>$ is a free quark state with momentum $p,\ \vert
\psi_p>$ is a dressed quark state, and $\vert p-k; k(\lambda,   c)>$ is a
state of a quark, of momentum $p-k,$ and a gluon of momentum $k,$
helicity $\lambda$ and color $c.$  We suppress quark color indices which
will appear in matrix form in what follows.  We label states by momenta
$p_+={1\over\sqrt{2}}\ (p_0+p_3), p_1,p_2,$  and $d^3k=dk_+d^2k=dk_+dk_1dk_2.$ Recall that in light-cone quantization momenta $P_+$ and
${\underline P}=(P_1,P_2)$ are kinematic while $P_-$ plays the role of a
Hamiltonian and generates evolution in the ``time'' variable
$x_+\ =\ {1\over \sqrt{2}}\ (x_0+x_3).$  For an on-shell zero mass
particle $p_-={\underline p}^2/2p_+.\   N$  in (1) is a normalization
factor.  Eq.(1) is illustrated in Fig.1.

\begin{center}
\begin{figure}
\epsfbox[0 0 213 73]{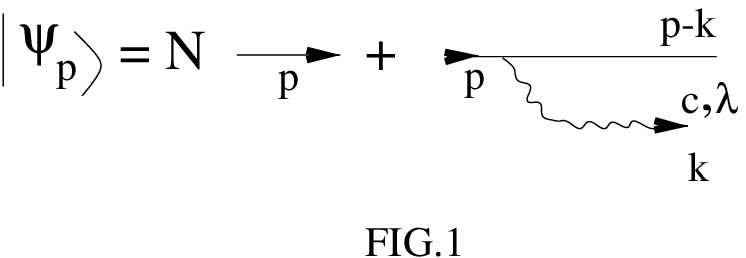}
\end{figure}
\end{center}

$\psi_\lambda^c$ is determined from light-cone perturbation theory to be

\begin{equation}
\delta^3(p-p^\prime)\psi_\lambda^c(k) = {<p^\prime-k;k(\lambda,c)\vert
H_I\vert p>\over (p^\prime-k)_-+k_- -p_-}
\end{equation}

\noindent with

\begin{equation}
 H_I=g\int d^3x {\bar q}(x) \gamma_\mu({\lambda^c\over 2}) q(x)A_\mu^c(x)
\end{equation}

\noindent where the gluon field is

\begin{equation}
A_\mu^c(x)= \sum_{\lambda = \pm}\int
{d^3k\over{\sqrt{(2\pi)^32k_+}}}[\epsilon_\mu^\lambda(k)a_\lambda^c(k) e^{i{\underline
k}\cdot{\underline x}-ik_+x_--ik_-x_+}+ h.c.]
\end{equation}

\noindent with

\begin{equation}
[a_{\lambda^\prime}^{c^\prime}(k^\prime), a_\lambda^{\dagger
c}(k)]=\delta_{\lambda^\prime\lambda}\delta_{c^\prime c}\delta^3(k^\prime-k).
\end{equation}

\noindent In (2) and (5) $\delta^3(p) = \delta(p_+) \delta^2({\underline p})$ while
$d^3x=dx_-d^2{\underline x}$ in (3).  It is useful to imagine the calculation being done
in a frame where $p_+$ is large and ${\underline p}=0$ in which case 

\begin{equation}
k_-={{\underline k}^2\over 2k_+},\ \ \ \ \ \  (p-k)_-={{\underline k}^2\over 2(p-k)_+}.
\end{equation}

\noindent In the soft gluon approximation $k_+/p_+ << 1$ and thus $k_->> (p-k)_-$ so
that only $k_-$ need be kept in the denominator in (2).  In addition, in light-cone
gauge, $A_+=0$, the polarization vectors can be written as

\begin{equation}
\epsilon_\mu^\lambda(k) = (\epsilon_+^\lambda, \epsilon_-^\lambda,{\underline
\epsilon}^\lambda) = (0, {{\underline \epsilon}^\lambda\cdot{\underline k}\over k_+},
{\underline \epsilon}^\lambda)
\end{equation}

\noindent and, because of the $1/k_+$ term, only $\epsilon_-^\lambda$ need be kept in
(2) in the soft gluon approximation.  Using (3)-(7) in (2) one finds

\begin{equation}
\psi_\lambda^c(k) = ({\lambda^c\over 2}) 2g {({\underline
\epsilon}^\lambda)^\ast\cdot{\underline k}\over {\underline k}^2} \ {1\over
{\sqrt{(2\pi)^32k_+}}}
\end{equation}

\vskip5pt
\noindent {\bf Problem 1(E)}:  Using the formula ${\bar
U}(p-k)\gamma_\mu {\bar U}(p)\approx 2p_+g_{\mu_-}$ for high
momentum Dirac spinors derive (8).

\section{Partons}

Define the gluon distribution of a state $\vert S(p)>$ by

\begin{equation}
xG_S(x,Q^2) = \sum_{\lambda,c}\int d^3kx\delta(x-k_+/p_+)\Theta(Q^2-{\underline
k}^2)<S(p)\vert a_\lambda^{c\dagger}(k)a_\lambda^c
(k)\vert S(p)>.
\end{equation}

\noindent The meaning of $xG_S$ is clear.  $xG_S(x,Q^2) dx$ is the number of gluons,
having longitudinal momentum between $xp_+$ and $(x + dx) p_+,$ localized in transverse
coordinate space to a region $\Delta x_\perp \sim 1/Q,$ in the state $\vert S(p)>.$  For
a quark, at order $g^2,$ one finds from (1)

\begin{equation}
xG_q(x,Q^2) = \sum_{\lambda,c}\int d^3k x \delta(x-k_+/p_+)\Theta(Q^2-{\underline
k}^2)\psi_\lambda^{c\dagger}(k)\psi_\lambda^c(k).
\end{equation}

\noindent Using (8) in (10) one finds

\begin{equation}
xG_q(x,Q^2) = \sum_c{\lambda^c\over 2}\ {\lambda^c\over 2}\int {4g^2\over (2\pi)^3}\
{d^2k\over {\underline k}^2}\ {dk_+\over 2k_+}\ x
\delta(x-k_+/p_+)\Theta(Q^2-{\underline k}^2).
\end{equation}

\noindent Using\ $\Sigma_c\ {\lambda^c\over 2}\ {\lambda^c\over 2} =
C_F={N_c^2-1\over 2N_c}$ and introducing an infrared cutoff, $\mu,$ for the
transverse momentum integral in (11) gives

\begin{equation}
x G_q(x,Q^2) = {\alpha C_F\over \pi} \ell n (Q^2/\mu^2).
\end{equation}

\noindent We note that if $xG(x,Q^2) = 3x G_q(x,Q^2)$ is taken one obtains a result for
the proton which is not unreasonable phenomenologically for $x \sim 10^{-2}-10^{-1}$ and
moderate $Q^2$ if $\mu$ is taken to be 100 MeV.

\section{Classical Fields}

One can associate a classical field with gluons in the quark.

\begin{equation}
A_i^{(c\ell)}(x) = \int d^3p^\prime<\psi_{p^\prime}\vert A_i^c(x)\vert \psi_p>.
\end{equation}

\noindent Using (1) and (8) in (13) one finds

\begin{equation}
A_i^{c(c\ell)}(x)=\int{d^3k\over (2\pi)^3} e^{-ik\cdot x}({\lambda^c\over 2}) {gk_i\over
{\underline k}^2k_+}
\end{equation}

\noindent or

\begin{equation}
A_i^{c(c\ell)}(x) = \int {d^3k\over (2\pi)^3} e^{-ik\cdot x}A_i^{c(c\ell)}(k)
\end{equation}

\noindent with
\begin{equation}
A_i^{c(c\ell)}(k) = {\lambda^c\over 2}\ {gk_i\over k_+}.
\end{equation}

\noindent In (14) the $k_+$ integration goes from $- \infty$ to $+ \infty.$  The region
$k_+ > 0$ comes from $\vert \psi_p>$ consisting of a bare quark and a gluon of momentum
$k$ while the region $k_+< 0$ comes from $\vert \psi_{p^\prime}>$ consisting of a bare
quark and a gluon.
\vskip5pt

\noindent{\bf Problem 2(E)}:  Take ${1\over k_+} = {1\over k_+-i\epsilon}$ in (16) and
show that

\begin{displaymath}
A_i^{c(c\ell)}(x) = - g({\lambda^2\over 2}){x_i\over 2\pi{\underline x}^2} \Theta(-x_-)\ 
{\rm and}\  F_{+_i}^{c(c\ell)}= {\partial\over \partial x_-} A_i^{c(c\ell)}=g({\lambda^c\over
2}) {x_i\over 2\pi{\underline x}^2} \delta(x_-).
\end{displaymath}

\vskip5pt
\noindent{\bf Problem 3(E)}:  Show that $G_q,$ as given in (10), can also be written as

\begin{displaymath}
xG_q(x,Q^2) = \int {d^3k\over (2\pi)^3} \delta(x_-k_+/p_+)\Theta(Q^2-{\underline
k}^2)\sum_{i,c}[A_i^{c(c\ell)}(k)]^2 2k_+.
\end{displaymath}

\section{Why Light-Cone Gauge is Special}

In order to understand why light-cone gauge plays a special role in describing high-energy hadronic states a simple calculation of the
near forward high-energy elastic amplitude for electron-electron scattering in QED is useful.  The graph to be calculated
is shown in Fig.2 and we imagine the calculation being done in the center of mass frame
with  p  being a right-mover ($p_+$large) and $p_1$ being a left-mover ($p_{1_-}$
large).  First we shall do the calculation in covariant gauge and afterwards in
light-cone gauge.

\begin{center}
\begin{figure}
\epsfbox[0 0 81 122]{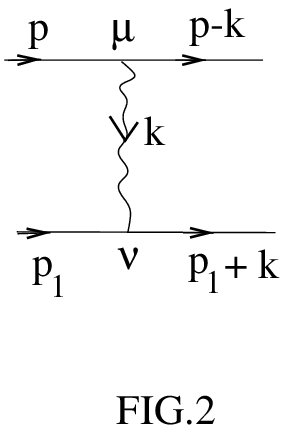}
\end{figure}
\end{center}

In covariant gauge the photon propagator, the k-line in Fig.2, is $D_{\mu\nu} = {-i\over
k^2} g_{\mu\nu}$ and the dominant term comes from taking $g_{\mu\nu} \to g_{- +}=1$
giving

\begin{equation}
T = ig{{\bar U}(p-k)\gamma_+ U(p)\over 2p_+}\ ig \ {{\bar U}(p_1+k) \gamma_-U(p_1)\over
2p_{1-}}\ {-i\over k^2}.
\end{equation}

\noindent Using ${\bar U}(p-k) \gamma_+U(p)\approx 2p_+$ and ${\bar
U}(p_1+k)\gamma_-U(p_1) \approx 2 p_{1-}$

\noindent one finds

\begin{equation}
T = {-ig^2\over {\underline k}^2}
\end{equation}

\noindent where we have used $k^2=2k_+ k_ - -{\underline k}^2\approx -{\underline k}^2$
since both $k_+$ and $k_-$ are required to be small from the mass shell conditions
$(p-k)^2 = (p_1+k)^2=0$ for the zero mass electrons.

Now suppose we do the calculation in light-cone gauge $A_+=0$ where the propagator is

\begin{equation}
D_{\mu\nu}(k) = {-i\over k^2}[g_{\mu\nu}-{\eta_\mu  k_\nu +\eta_\nu k_\mu\over \eta\cdot k}]
\end{equation}

\noindent where $\eta\cdot V = V_+$ for any vector $V_\mu.$  Now the dominant term comes
from taking $D_{-i}$ giving

\begin{equation}
T=iq{{\bar U}(p-k)\gamma_+U(p)\over 2p_+}\ ig\ {{\bar U}(p_1+k){\underline \gamma}\cdot
{\underline k}\  U(p_1)\over 2p_{1-}}\ {-i\over k_+k^2}.
\end{equation} 

\noindent Using ${\bar U}(p_1+k) {\underline \gamma}\cdot {\underline k}\  U(p_1) = {\underline k}^2$ one finds

\begin{equation}
T = {-ig^2\over 2p_{1-}k_+}.
\end{equation}

\noindent Now $(p_1+k)^2=0$\  gives\    $2p_{1-}k_+\approx {\underline k}^2$ so that (18) and (21) agree as expected.

The result (18) comes about in a natural way ${{\bar U}(p-k)\gamma_+U(p)\over 2p_+}$ and ${{\bar U}(p_1+k)\gamma_-U(p_1)\over
2p_{1-}}$ are the classical currents, equal to 1, of particles moving along the light-cone while the $1/{\underline k}^2$ factor is
just the (instantaneous) potential between the charges.  However, when one uses $A_+=0$ light-cone gauge the dominant part of the
current for left moving particles is forbidden and one must keep the small transverse current ${\underline k}/2p_{1-}.$  However, the
smallness of the current is compensated by the factor $1/k_+$ in the light-cone gauge propagator.  In coordinate space the $1/k_+$
comes about from the potential acting over distances $x_-\approx 1/k_+={2p_{1-}\over {\underline k}^2}$ so that the potential is very
non-local and non-causal.  By choice of the $i\epsilon$ prescription\cite{Yu,Kov} for $1/k_+$ one can put these non-causal
interactions completely before the scattering (initial state) or completely after the scattering (final state).  In Problem 2 the
potential $A_i$ exhibits this non-causal behavior with our choice of $i\epsilon$ placing the long time behavior in the initial state,
the $\Theta(-x_-)$ term.

\section{High Momentum Particles and\newline
 Wilson Lines$^{3-5}$}

Suppose a quark of momentum  $p$   is a high-energy right mover, that is $p_+>> p_-,{\underline p}.$  Then so long as one does not
choose to work in $A_-=0$ light-cone gauge the dominant coupling of gauge fields to $p$  are classical (eikonal) when  $p$ passes
some QCD hadron or source.  To be specific suppose  $p$ scatters on a hadron elastically and with a small momentum transfer.  We may
view the interactions as shown in Fig.3 for a three-gluon exchange term.  Of course to get the complete scattering one has to sum over
all numbers of gluon exchanges.  Call $S(p_+,{\underline b})$ the S-matrix for scattering of the right moving quark on the target. 
(We imagine that the target hadron has large gluon fields making it necessary to find a formula which includes all gluon
interactions.)  Although the right moving quark has sufficient momentum so that it moves close to the light-cone we do assume that the
momentum is not so large that higher gluonic components of the quark wavefunction need be considered.  Since the probability of extra
transverse gluons being present in the wavefunction is in general proportional to $\alpha y,$ with\  $y$\ being the quark rapidity, we
suppose $\alpha y << 1.$

The graph shown in Fig.3 can be written as

\begin{center}
\begin{figure}
\epsfbox[0 0 148 124]{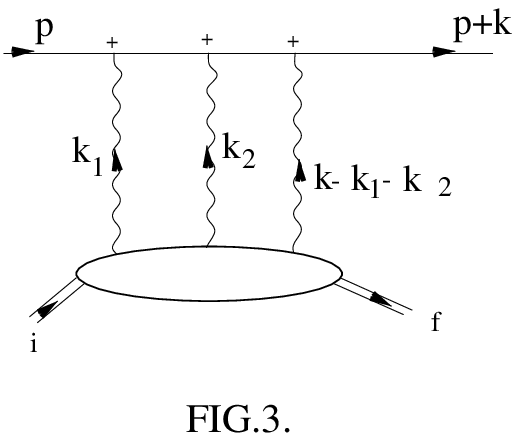}
\end{figure}
\end{center}

\begin{displaymath}
S(p_+,{\underline b})=\int \{{1\over{\sqrt{2(p+k)_+}}}{\bar
U}(p+k)igT^a\gamma_+{i\over\gamma\cdot (p+k_1+k_2)}igT^b\gamma_+{i\over\gamma\cdot(p+k_1)}
\end{displaymath}

\begin{equation}
\cdot  igT^c\gamma_+
U(p){1\over{\sqrt{(2p_+)}}}\}
{d^2k dk_+\over (2\pi)^3} e^{i{\underline k}\cdot {\underline b}}{d^4k_1d^4k_2\over (2\pi)^8} M_{abc}(k_1,k_2)
\end{equation} 

\noindent where

\begin{displaymath}
M_{abc}(k_1,k_2)=\int d^4x_1d^4x_2d^4x e^{i(k-k_1-k_2)\cdot x +ik_2\cdot x_2+ik_1\cdot x_1}
\end{displaymath}
\begin{equation}
\cdot <f\vert TA_+^a(x)A_+^b(x_2)A_+^c(x_1)\vert i>.
\end{equation}

\noindent It is straightforward to evaluate the $\{\ \}$-term in (22) and one gets

\begin{equation}
\{\ \} = igT^aigT^bigT^c{i\over k_{1-}+i\epsilon}\ {i\over (k_1+k_2)_-+ i\epsilon}.
\end{equation}

\noindent Now do the $d^2k_1 dk_{1+}$ and $d^2k_2dk_{2+}$ integrals followed by $d^2x_1dx_{1-}$ and $d^2x_2dx_{2-}.$  This sets
${\underline x}_1={\underline x}_2={\underline x}$ and $x_{1-} = x_{2-}=x_-.$  The only non-zero term in the time-ordered product is
proportional to $\Theta(x_- -x_{2-})\Theta(x_{2-}-x_{1-})$ and this factor along with the exponentials in $x_{1-}, x_{2-}$ and $x_-$
allow the $dk_{1-}$ and $dk_{2-}$ integrals to be done over the poles in (24).  We get finally

\begin{displaymath}
S_{fi}(p_+,{\underline b})=(ig)^3<f\vert\int_{-\infty}^\infty dx_+A_-({\underline b}, x_+)\int_{-\infty}^{x_+} dx_{1+}A_-({\underline
b}, x_{2+})
\end{displaymath}
\begin{equation}
\cdot \int_{-\infty}^{x_{2+}} dx_{1+}A_-({\underline b},x_{1+})\vert i>
\end{equation}

\noindent where $A_\mu=T^aA_\mu^a$ and we have suppressed the variable $x_-=0$ in the $A's$ in (25).

The general term is now apparent, and one has in the general case

\begin{equation}
S_{fi}(p_+,{\underline b}) = <f\vert P e^{{ig}\int_{-\infty}^\infty dx_+A_-({\underline b},x_+)}\vert i>
\end{equation}

\noindent where  $P$ denote an $x_+-$ ordering of the matrices $A$ where $A's$ having larger values of $x_+$ come to the left of
those having smaller values.

\vskip5pt
\noindent {\bf Problem 4(M)}:  Show that the time orderings different from $\Theta(x_+-x_{2+})\Theta(x_{2+}-x_{1+})$ give no
contribution to (22).

Finally, a word of caution in using (26).  In general there are singularities present when two values of $x_+$, in adjoining $A's,$
become equal, although in many simple models the $x_+-$ integrations are regular.  If singularities in the $x_+$ integrations arise it
is generally possible to extract the leading logarithmic contributions to the scattering amplitude by carefully examining the
singularities\cite{Bal}.  A detailed discussion of this is, however, far beyond the scope of these lectures.

\section{Dual Descriptions of Deep Inelastic Scattering; Bjorken and Dipole Frames}

Particular insight into the dynamics of a process often occurs by choosing a particular frame and an appropriate gauge.  Indeed the
physical picture of a process may change dramatically in different frames and in different gauges.  In a frame where the parton picture
of a hadron is manifest saturation shows up as a limit on the occupation number of quarks and gluons, however, in a different (dual)
frame saturation appears as the unitarity limit for scattering of a quark or of a gluon dipole at high-energy\cite{Mue}.

To see all this a bit more clearly consider inelastic lepton-nucleon scattering as illustrated in Fig.4 where a lepton emits a virtual
photon which then scatters on a nucleon.  We suppose $Q^2= - q^2$ is large.  The two structure functions, $F_1$ and $F_2,$ which
describe the cross section can depend on the invariants $Q^2$ and $x = {Q^2\over 2P\cdot q}.$  To make the parton picture manifest we
choose $A_+=0$ light-cone gauge along with the frame

\begin{center}
\begin{figure}
\epsfbox[0 0 230 169]{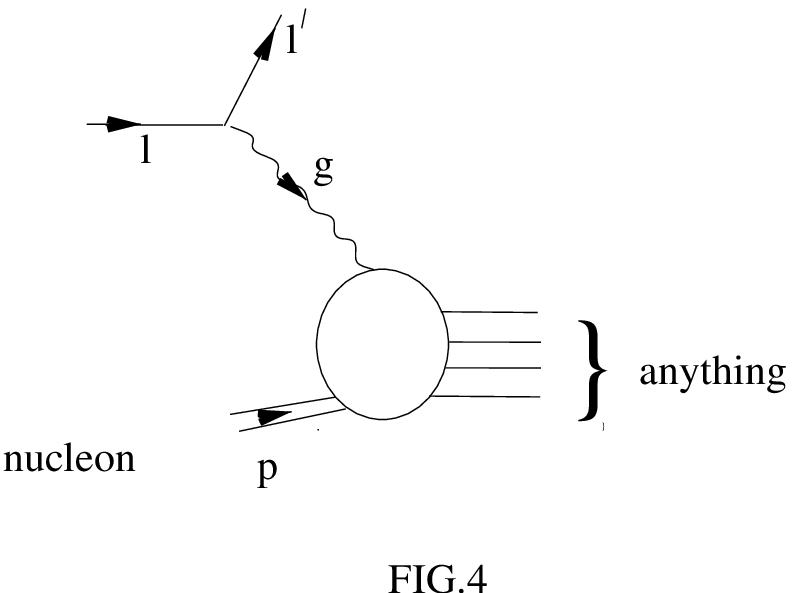}
\end{figure}
\end{center}

\begin{displaymath}
P = (P + {M^2\over 2P}, 0,0,P)
\end{displaymath}

\noindent and

\begin{displaymath}
q = (q_0, {\underline q}, q_z=0),
\end{displaymath}

\noindent and where $P \rightarrow \infty.$  This is the Bjorken frame.  We note that $q_0={P\cdot q\over P}$ goes to zero as $P
\rightarrow \infty$ so that the virtual photon momentum is mainly transverse to the nucleon direction.  This last fact means that the
virtual photon is a good analyzer of transverse structure since it is absorbed over a transverse distance $\Delta x_\perp\sim 1/Q.$ 
Since $\Delta x_\perp$ is very small at large \ $Q$\ the virtual photon is absorbed by, and measures, individual quarks.

\vskip5pt
\noindent{\bf Problem 5(H)}:  Use the uncertainty principle to show that the time, $\Delta\tau,$ over which $\gamma^\ast(q)$ is
absorbed by a quark is $\Delta\tau \approx 2xP/Q^2.$  You may assume that ${\underline k}^2/Q^2 << 1$ where the process of absorption
of the photon is illustrated in Fig.5. 

\begin{center}
\begin{figure}
\epsfbox[0 0 130 173]{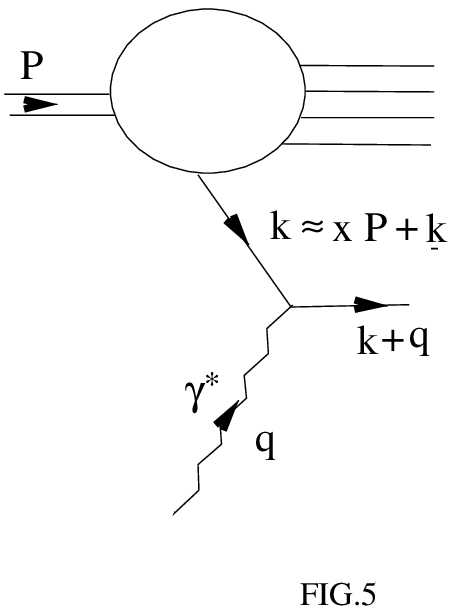}
\end{figure}
\end{center} 

The result of problem 5 and of the result that $\Delta x_\perp \sim 1/Q$ motivates the formula

\begin{equation}
F_2(x,Q^2) = \sum_f e_f^2[x q_f(x,Q^2) + x {\bar q}_f(x,Q^2)]
\end{equation}

\noindent which says that the structure function $F_2$ is proportional to the charge squared of the quark, having flavor $f,$ absorbing
the photon and proportional to the number of quarks having longitudinal momentum fraction x and localized in transverse coordinate
space to a size $1/Q.\   F_2$  is given in terms of the longitudinal and transverse virtual photon cross sections on the proton as

\begin{equation}
F_2 = {1\over 4\pi^2\alpha_{em}} Q^2[\sigma_T + \sigma_ L].
\end{equation}

\noindent Eq.(27) is the QCD improved parton model.  In more technical terms $Q^2$ is a renormalization point which in the parton
picture is a cutoff of the type given in (9) for gluon distributions and here occurring for quark distributions.

Now consider the same process in the dipole frame pictured in Fig.6 where

\begin{center}
\begin{figure}
\epsfbox[0 0 173 162]{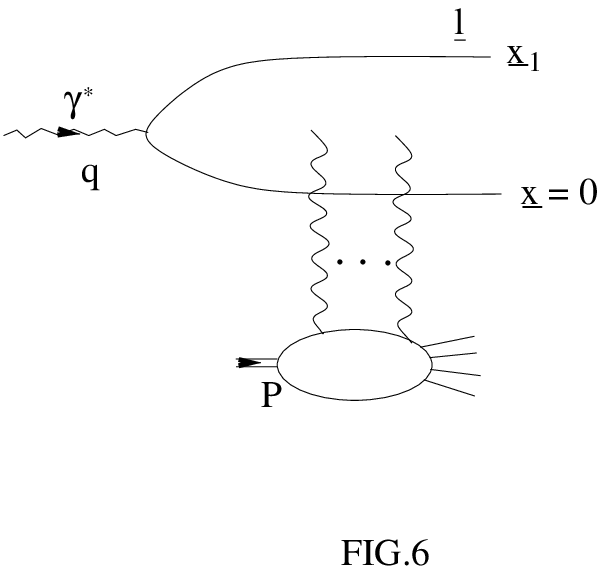}
\end{figure}
\end{center}

\begin{displaymath}
P=(P+{M^2\over 2P}, 0,0,P)
\end{displaymath}

\begin{displaymath}
q=({\sqrt{q^2-Q^2}},0,0,-q)
\end{displaymath}

\noindent and where $q/Q>>1$ but where $q$ is fixed as $x$ becomes small so that most of the energy in a small-x process, and it is
only for small-x scattering that the dipole frame is useful, still is carried by the proton.  Now we suppose that a gauge different
from $A_+=0$ is being used, for example a covariant gauge or the gauge $A_-=0.$  In the dipole frame the process looks like
$\gamma^\ast\rightarrow$ ${\rm quark-\overline{quark}}$ followed by the scattering of the quark-antiquark dipole on the
nucleon\cite{Bjo,ler,Fra}.  The splitting of $\gamma^\ast$ into the quark-antiquark pair is given by lowest order perturbation
theory while all the dynamics is in the dipole-nucleon scattering.  In this frame the partonic structure of the nucleon is no longer
manifest, and the virtual photon no longer acts as a probe of the nucleon.

Equating these two pictures and fixing the transverse momentum of the leading quark or antiquark one has\cite{Mue}

\begin{displaymath}
e_f^2{d(xq_f+x{\bar q}_f)\over d^2bd^2{\underline \ell}} = {Q^2\over 4\pi^2\alpha_{em}} \int {d^2x_1d^2x_2\over 4\pi^2}\int_0^1
dz{1\over 2} \sum_\lambda\psi_{T\lambda}^{f\ast}({\underline x}_2,z,Q)\psi_{T\lambda}^f({\underline x}_1,z,Q)
\end{displaymath}
\begin{equation}
\cdot e^{-i{\underline \ell}\cdot({\underline x}_1-{\underline x}_1)} [S^\dagger({\underline x}_2)S({\underline x}_1) -
S^\dagger({\underline x}_2)- S({\underline x}_1) + 1],
\end{equation}

\noindent where the $\gamma^\ast$ wavefunction is

\begin{equation}
\psi_{T\lambda}^f({\underline x},z,Q) = \{{\alpha_{em}N_c\over 2\pi^2}z(1-z)[z^2+(1-z)^2]
Q^2\}^{1/2} e_f K_1({\sqrt{Q^2{\underline x}^2z(1-z)}}){{\underline \epsilon}^\lambda\cdot{\underline x}\over\vert {\underline
x}\vert}.
\end{equation}

\noindent We have written (29) for a fixed impact parameter, ${\underline b},$ which is a (suppressed) variable in S.  To make the
identification exhibited in (29) requires that the struck quark shown in Fig.5 not have final state interactions, and this requires the
special choice of $i\epsilon's$ in the light-cone gauge, $A_+=0,$ used to identify the quark distributions.  With this choice of
$i\epsilon's$ the lefthand side of (29) refers to the density of quarks in the nucleon wavefunction while the righthand side of (29)
refers to  the transverse momentum spectrum of quark jets produced.  This identification is only possible when final state
interactions are absent.  Finally, it is not hard to see the origin of the final factor on the righthand side of (29).  This is just
$[S^\dagger({\underline x}_2)-1][S({\underline x}_1)-1],$  the product of the $T-$ matrices for the dipole ${\underline x}_1$ in the
amplitude and the dipole ${\underline x}_2$ in the complex conjugate amplitude.

\section{Quark Distributions in a large Nucleus; Quark Saturation at One-Loop}

In general it is very difficult to evaluate the $S-$ matrices appearing in (29).  However,  if the target is a large nucleus one can
define an interesting model, if not a realistic calculation for real nuclei, by limiting the dipole nucleon interaction to one and two
gluon exchanges.  That is, we suppose the dipole-nucleon interaction is weak.  Despite this assumed weakness of interaction the dipole
nucleus scattering can be very strong if the nucleus is large enough.  We begin with the term $S({\underline x}_1)$ in (29), an elastic
scattering of the dipole on the nucleus in the amplitude with no interaction at all in the complex conjugate amplitude.  This term is
illustrated in Fig.7.

\begin{center}
\begin{figure}
\epsfbox[0 0 191 123]{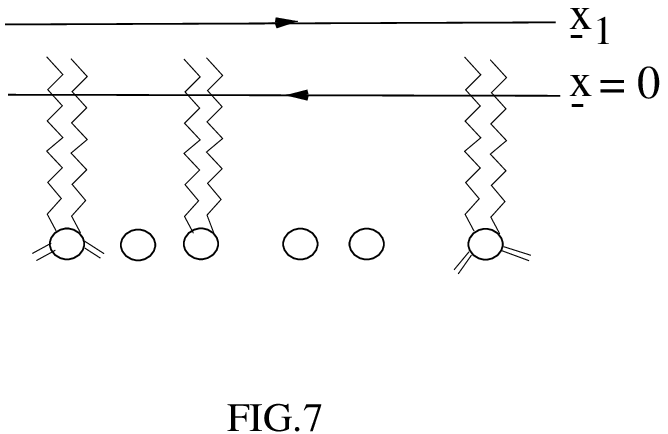}
\end{figure} 
\end{center}

Now $\vert S({\underline x}_1)\vert^2$ is the probability that the dipole does not have an inelastic interaction as it passes through
the nucleus.  We can write

\begin{equation}
\vert S({\underline x}_1)\vert^2 = e^{-L/\lambda}
\end{equation}

\noindent where $L = 2{\sqrt{R^2-{\underline b}^2}}$ is the length of nuclear matter that the dipole traverses at impact parameter
${\underline b}$ for a uniform spherical nucleus of radius $R,$ and $\lambda$ is the mean free path for inelastic dipole-nucleon
interactions.  Using

\begin{equation}
\lambda = [\rho \sigma]^{-1}
\end{equation}

\noindent with $\rho$ the nuclear density one has

\begin{equation}
S({\underline x}_1, {\underline b}) = e^{-2{\sqrt{R^2-b^2}}\rho\sigma({\underline x}_1)/2}
\end{equation}

\noindent if we suppose $S$ is purely real.  Detailed calculation gives\cite{tel}

\begin{equation}
\sigma({\underline x}) = {\pi^2\alpha\over N_c} x G(x, 1/{\underline x}^2){\underline x}^2.
\end{equation}
\vskip5pt

\noindent{\bf Problem 6(H$^\ast$)}:  Check (34) for scattering of a dipole on a bare quark.

Thus we know how to calculate the last three terms in [\ ]  in (29).  What about the $S^\dagger S$ term?  Graphically this term is
illustrated in Fig.8 for some typical elastic and inelastic interactions of the dipoles with the nucleons in the nucleus.  Let us
focus on the final interaction, the one closest to the cut (the vertical line) in Fig.8.  One can check that interactions with the
${\underline x} = {\underline 0}$ line cancel between interactions in the amplitude and the complex conjugate amplitude.

\begin{center}
\begin{figure}
\epsfbox[0 0 291 162]{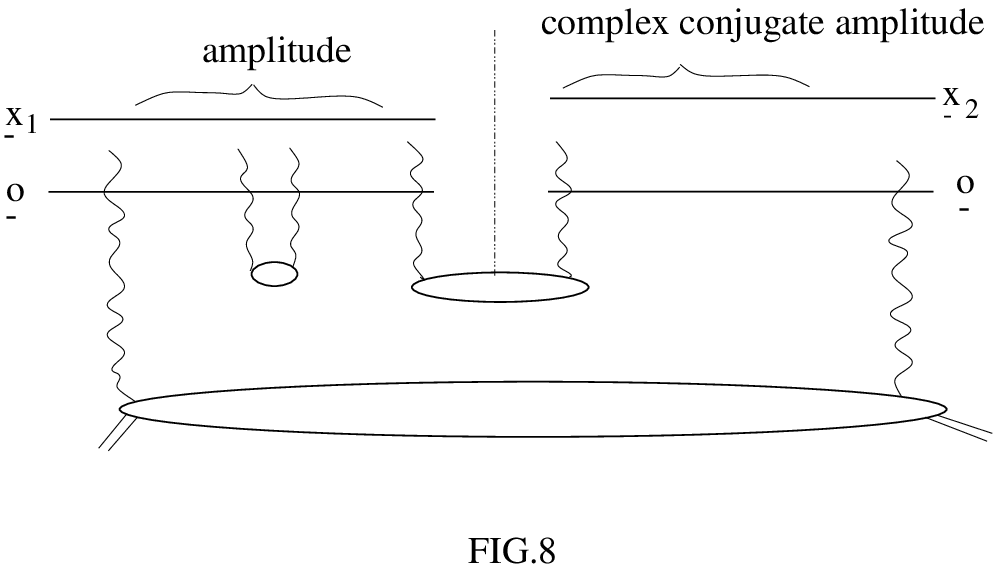}
\end{figure}
\end{center}

\vskip5pt
\noindent{\bf Problem 7(H)}:  Verify, using $S^\dagger S = 1,$ that the three interactions shown in Fig.9 cancel. You may assume that
$T$ is purely imaginary $(S= 1 - i T)$ although this is not necessary for the result.

\begin{center}
\begin{figure}
\epsfbox[0 0 325 127]{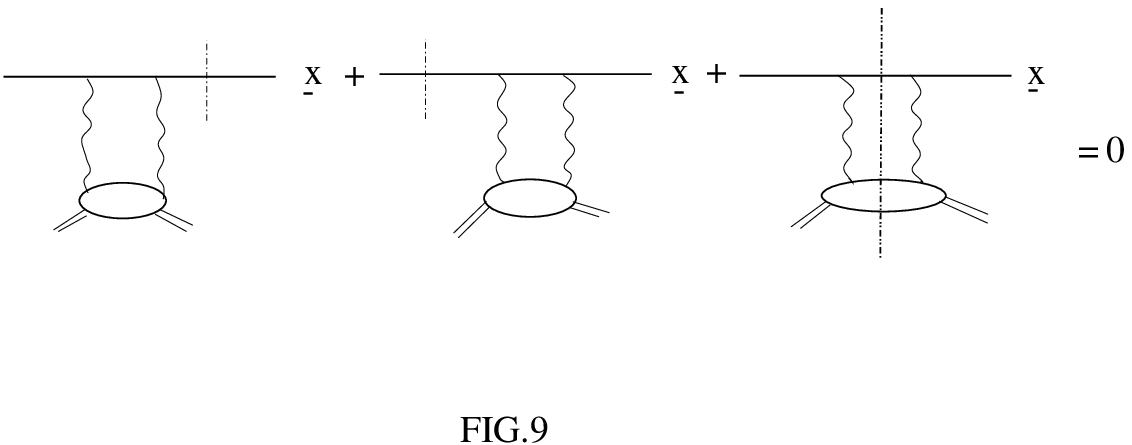}
\end{figure}
\end{center}

\noindent{\bf Problem 8(H)}:  Again, using $S^\dagger S=1,$ show that the two interactions shown in Fig.10 cancel.  Assume  $T$ is
purely imaginary.

\begin{center}
\begin{figure}
\epsfbox[0 0 207 148]{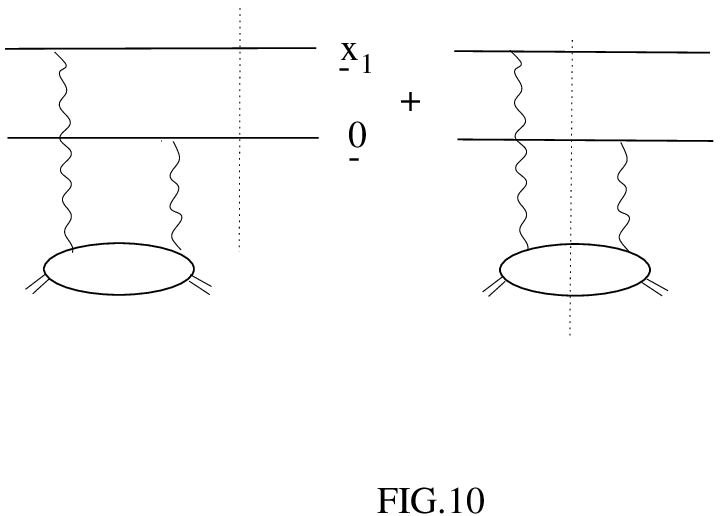}
\end{figure}
\end{center}

The results of problems 7 and 8 establish that one need not consider interactions with the ${\underline x}=0$ line in evaluating the
$S^\dagger({\underline x}_2) S({\underline x}_1)$ term in (29).  Thus, we are left only with interactions on the ${\underline
x}_1-$line in the amplitude andwith the ${\underline x}_2$-line in the complex conjugate amplitude.  The result of problem 8 allows
one to transfer the interactions with the ${\underline x}_2$-line in   the complex conjugate amplitude to interactions with a line
placed at ${\underline x}_2$ in the amplitude\cite{Zak}, although this does require that $S$ be real.  This is illustrated in the equality between the terms in Fig.11a and Fig.11b and reads

\begin{center}
\begin{figure}
\epsfbox[0 0 153 119]{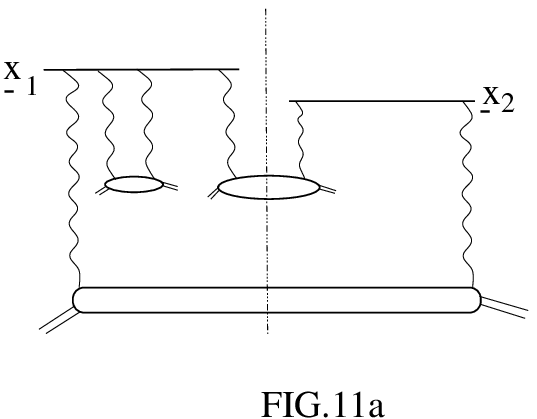}
\end{figure}
\end{center}
\begin{center}
\begin{figure}
\epsfbox[0 0 179 147]{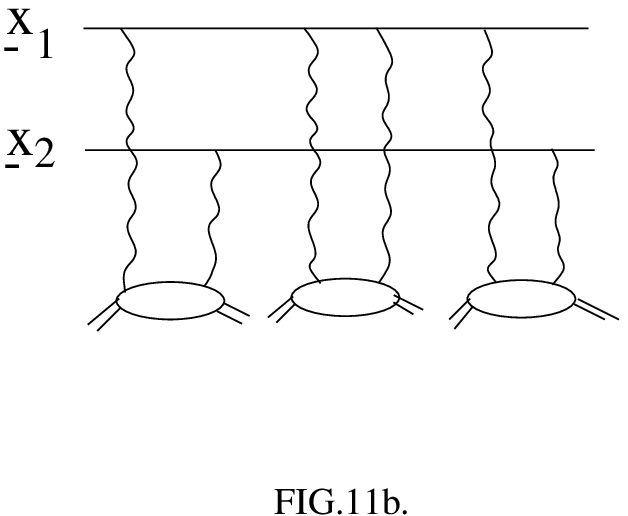}
\end{figure}
\end{center}

\begin{equation}
S^\dagger({\underline x}_2) S({\underline x}_1) = S({\underline x}_1-{\underline x}_2).
\end{equation}

\noindent Eq.(35) is a beautiful result, but unfortunately it likely has a limited validity.  It appears to require a maximum of two
gluon lines exchanged between any given nucleon and the dipole lines as well as the requirement of a purely real S-matrix for elastic
scattering.  Thus using (29), (33) and (35) we arrive at\cite{Mue}

\begin{displaymath}
e_f^2{d(xq_f+x{\bar q}_f)_A\over d^2\ell d^2b}= {Q^2\over 4\pi^2\alpha_{em}}\int {d^2x_1d^2x_2\over 4\pi^2}\int_0^1 dz {1\over
2}\sum_\lambda \psi_{T\lambda}^{f\ast}\psi_{T\lambda}^f e^{-i{\underline\ell}\cdot({\underline x}_1-{\underline x}_2)}\cdot
\end{displaymath} 
 \begin{equation}
\cdot[1 + e^{-({\underline x}_1-{\underline x}_2)^2{\bar Q}_S^2/4}-e^{-{\underline x}_1^2{\bar Q}_S^2/4}-e^{-{\underline x}_2^2{\bar
Q}_S^2/4}]
\end{equation}

\noindent with

\begin{equation}
{\bar Q}_S^2 = {C_F\over N_c} Q_S^2
\end{equation}

\noindent and  

\begin{equation}
Q_S^2={8\pi^2\alpha N_c\over N_c^2-1}\ {\sqrt{R^2-b^2}}\ \rho x G,
\end{equation}

\noindent where ${\bar Q}_S$ is the quark saturation momentum and $Q_S$ is the gluon saturation momentum.  In (38) $xG$ is the gluon
distribution in a nucleon.  Eq.(36) is a complete solution to the sea quark distribution of a nucleus in the one-quark-loop
approximation.

\vskip5pt
\noindent{\bf Problem 9(M-H)}:  Use (30) and (36) to show that$\cite{Mue}$

\begin{equation}
{dx(q_f+{\bar q}_f)A\over d^2bd^2\ell} = {N_c\over 2\pi^4}
\end{equation}

\noindent when ${\underline \ell}^2/{\bar Q}_S^2 << 1.$ 

Eq.(39) gives meaning to the idea of saturation as a maximum occupation number
for, in this case, quarks.  As the nucleus gets larger and larger, that is as  R  grows, ${\bar Q}_S^2$ grows and so there are more and
more sea quarks in the nuclear wavefunction, nevertheless, the 2-dimensional occupation number hits a constant upper bound for momenta
below the saturation momentum.

\vskip5pt
\noindent{\bf Problem 10(M-H)}:  Show that $1/2$ of (39) comes from the first term in [\ ] in (36) and that $1/2$ comes from the
second term with the third and fourth term being small. The second term can be viewed as due to inelastic reactions while the first
term is the elastic shadow of these inelastic reactions of a dipole passing over the nucleus.

\section{Gluon Saturation in a Large Nucleus; the McLerran Venugopalan Model$^{12}$}

In order to directly probe gluon densities it is useful to introduce the ``current''

\begin{equation}
j= - {1\over 4} F_{\mu\nu}^a F_{\mu\nu}^a.
\end{equation}

\noindent We shall then calculate the process $j + A \to$ gluon $({\underline \ell}) +$ anything.  I shall interpret the process in a
slightly modified Bjorken frame, and in light-cone gauge, while we shall do the calculation in a covariant gauge and in the rest system
of the nucleus\cite{Kov}.

For the interpretation we take the momenta of a nucleon in the nucleus and the current to be

\begin{displaymath}
p=(p+{M^2\over 2p}, 0,0,p)
\end{displaymath}
\begin{displaymath}
q=(0,0,0,-2xp).
\end{displaymath}

\noindent  For small $x\  p$ is large since $p = {Q\over 2x}$ and this frame is much like
an infinite momentum frame.  If we choose an appropriate light-cone gauge, one that eliminates final state interactions, then the
transverse momentum and the x distribution of gluons in the nuclear wavefunction is the same as the distribution of produced gluon jets
labeled by $\ell$ in Fig.12.

\begin{center}
\begin{figure}
\epsfbox[0 0 165 146]{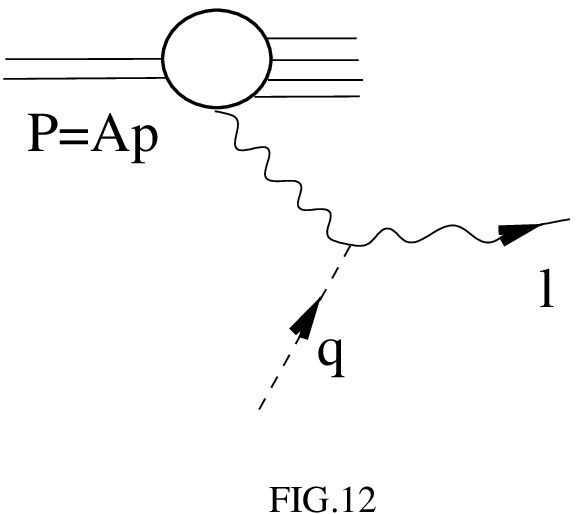}
\end{figure}
\end{center}

In order to do the calculation of the spectrum of produced gluon jets we carry out a multiple scattering calculation, a term of which
is illustrated in Fig.13\cite{Kov}.  The calculation is simplest to do in covariant gauge.  Then

\begin{equation}
{dxG_A\over d^2bd^2\ell} = \int_0^{2{\sqrt{R^2-b^2}}} dz \rho xG(x, 1/{\underline x}^2) e^{-{z\over  L}{\underline
x}^2Q_S^2/4}e^{-i{\underline \ell}\cdot {\underline x}}{d^2x\over 4\pi^2}
\end{equation}

\noindent where $xG$ is the gluon distribution for a single nucleon, and we make use of the coordinate space interpretation where it
corresponds to the gluon in the complex conjugate amplitude being separated from that in the amplitude by $\Delta x_\perp \sim 1/Q.$ 
We note that the momentum space distribution of gluon produced off a single nucleon, the unintegrated gluon distribution, is given by

\begin{center}
\begin{figure}
\epsfbox[0 0 203 170]{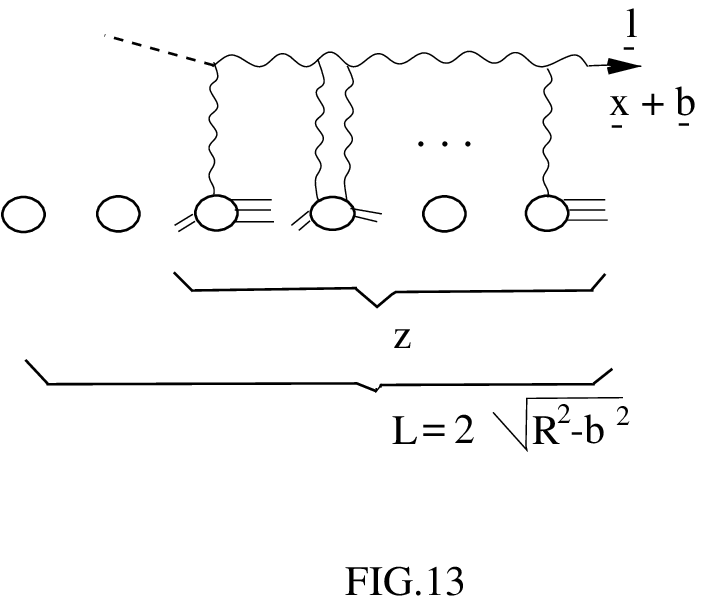}
\end{figure}
\end{center}

\begin{equation}
{dxG\over d^2\ell} = \int {d^2x\over 4\pi^2} xG(x, 1/{\underline x}^2) e^{-i{\underline \ell}\cdot{\underline x}}
\end{equation}

\noindent as can be easily verified by integrating the lefthand side of (42) over $d^2{\underline \ell}\Theta(Q^2-{\underline
\ell}^2).$  Carrying out the $z-$integration in (4) one gets\cite{Jal}

\begin{equation}
{dxG_A\over d^2bd^2\ell} = {N_c^2-1\over 4\pi^4\alpha N_c}\int d^2x{1-e^{-{\underline x}^2Q_S^2/4}\over {\underline x}^2}
e^{-i{\underline \ell}\cdot{\underline x}}
\end{equation}

\noindent where $Q_S^2$ is given in (38).  We also note that the $e^{-{z\over L}{\underline x}^2Q_S^2/4}$ factor in (41) is just
$S^\dagger({\underline b})S({\underline b}+{\underline x}) = S({\underline x})$ for a gluon dipole to pass over a length $z$  of
nuclear material.  Thus the derivation of (41) closely resembles that leading to (35) but now for gluons rather than quarks.  Finally,
when ${\underline \ell}^2/Q_S^2 <<1$ we find from (43)

\begin{equation}
{dxG_A\over d^2bd^2\ell}\  _{_{\longrightarrow}\atop{{\underline \ell}^2/Q_s^2<<1}}\ {N_c^2-1\over 4\pi^3\alpha N_c} \ell
n(Q_S^2/{\underline \ell}^2)
\end{equation}

\noindent while for ${\underline \ell}^2/Q_S^2 >> 1$ Eq.(43) gives the nuclear gluon distribution as simply a factor $A$ times the
nucleon gluon distribution.  Eq.(44) shows that saturation is somewhat more complicated for gluons.  The factor of $N_c^2-1$ counting
the number of species of gluons is expected as is the factor of $\alpha N_c$ in the denominator.  What is a little surprising is the
log factor for which there is not yet a good intuitive understanding.  Whether this log is a general factor or a peculiarity of the
present model is not known for sure\cite{Mue,Ian}.

\section{The Golec-Biernat W\"usthoff Model$^{15}$}

We turn for a while to some phenomenology to see whether there is evidence for saturation of parton densities in deep inelastic
lepton-proton scattering.  So far the best way that has been found to approach this problem is through a simple model of deep
inelastic and diffractive scattering inspired by the idea of saturation.  We can motivate this discussion by going back to (36) and,
supposing that such a picture might apply to a proton as well as a large nucleus, summing over  \ $f$\ and integrating over
${\underline \ell}$ and ${\underline b}$ obtain

\begin{equation}
F_2(x,Q^2) = {Q^2\over 4\pi^2\alpha_{em}}\int d^2x \int_0^1 dz \sum_\lambda\vert \psi_{T\lambda}^f({\underline x},
z,Q)\vert^2\int d^2b[1-e^{-{\underline x}^2{\bar Q}_S^2/4}].
\end{equation}

\noindent Now our ${\bar Q}_S^2$ naturally depends on the impact parameter $b,$ as indicated in (37) and (38) for a nuclear target,
however, as an approximation we suppose

\begin{equation}
\int d^2b[1-e^{-{\underline x}^2{\bar Q}_S^2/4}]=\sigma_0(1-e^{-{\underline x}^2/4R_0^2})
\end{equation}

\noindent where $R_0$ will now be taken to depend only on $x.$  Thus

\begin{equation}
F_2(x,Q^2) = {Q^2\over 4\pi^2\alpha_{em}}\int d^2x \int_0^1 dz \sum_\lambda\vert \psi_{T\lambda}^f({\underline x}, z,
Q)\vert^2(1-e^{-{\underline x}^2/4R_0^2})\sigma_0
\end{equation}

\noindent which is the formula used by Golec-Biernat and W\"usthoff\cite{Gol}.  In addition it is then natural to take the
diffractive cross section to be given by the shadow of the inelastic collisions in which case one replaces $\sigma_0(1-e^{{\underline
x}^2/4R_0^2}) = \sigma_0(1-S)$ by ${1\over 2} \sigma_0(1-S)^2$ giving

\begin{equation}
F_2^{\scriptscriptstyle D}(x,Q^2) = {Q^2\over 4\pi^2\alpha_{em}}\int d^2x \int_0^1 dz \sum_\lambda \vert
\psi_{T\lambda\lambda}^f({\underline x}, z,Q)\vert^2 {1\over 2}\sigma_0(1-e^{-{\underline x}^2/4R_0^2})^2.
\end{equation}

\noindent Eq.(47) represents a total cross section, (48) represents an ``elastic'' cross section while the inelastic contribution would
have a factor of ${1\over 2} \sigma_0(1-e^{-{\underline x}^2/2R_0^2})$ replacing the last factors in (47) and (48).

Golec-Biernat and W\"usthoff include a quark-antiquark-gluon scattering term in addition to the quark-antiquark dipole
term given by (48) so that larger mass diffractive states can also be described.  For our purposes this is a detail which in any
case introduces no new parameters.  The model has three parameters

\begin{equation}
\sigma_0 = 23mb, \ \ R_0^{-2} = <{\bar Q}_S^2> = ({x_0\over x})^\lambda GeV^2
\end{equation}

\noindent where $\lambda = 0.3$ and $x_0=3${\rm x}$10^{-4}.$  With these three parameters a good fit to low and moderate $Q^2$ and low
$x\  F_2$ and $F_2^{\scriptstyle D}$ data is obtained.  In fact the fit is surprisingly good over a range of $Q^2$ which is remarkably
large given that there is no QCD evolution present in the model.  We shall have to wait for further tests and refinements to be sure
that the fits are meaningful, but we may have the first bit of evidence for saturation effects.  The fact that $<{\bar Q}_S^2>$ is in
the $1 GeV^2$ region is reasonable.

\section{Measuring Dipole Cross Sections}

Refer back to (47).  It is easy to check that

\begin{equation}
F_2 = c \int_{1/Q^2}^{1/Q_S^2}\ {d{\underline x}^2\over {\underline x}^2}
\end{equation}

\noindent when $Q^2/Q_S^2 >>1.$  Thus, although we may view $F_2$ as being given by a dipole cross section the size of the dipole is
not well determined by $Q^2$ but, rather, varies between $1/Q$ and $1/Q_S.$  Thus currently deep inelastic structure functions are not
well suited for determining dipole cross sections. The situation should improve considerably when the longitudinal structure function
is measured, for in this case the dipole size will be fixed to be of size $1/Q$ and will give a direct measure of the dipole  cross
section.

\vskip5pt
\noindent{\bf Problem 11(E)}:  Use 30 and (45) to derive (50).  What is c?

At present the best place to measure dipole cross sections, and thus to see how close or how far one is from finding unitarity limits,
appears to be in the production of longitudinally polarized vector mesons\cite{Mun}.

The cross section for $\gamma_ L^\ast+$ proton $\rightarrow \rho_ L +$ proton can be written as

\begin{equation}
{d\sigma^{\gamma_ L^\ast\rightarrow\rho}\over dt} = {1\over 4\pi}\vert\int d^2xd^2b\int_0^1 dz \psi_\rho^\ast({\underline x}, z)
(1-S({\underline x}, {\underline b}))\psi_{\gamma^\ast}({\underline x},z,Q)^{i{\underline b}\cdot{\underline \Delta}}\big\vert^2
\end{equation}

\noindent where $t = - {\underline \Delta}^2$ is the momentum transfer.  The process is pictured in Fig.14 and can be viewed in three
steps. (i)  The virtual photon breaks up into a quark-antiquark dipole of size ${\underline x}$ with  $z$  being the longitudinal
momentum fraction of the $\gamma^*$ carried by either the quark or antiquark.(ii) The dipole scatters elastically on the proton with 
scattering amplitude $1-S({\underline x}, {\underline b})$ where ${\underline x}$ is the dipole size and ${\underline b}$ the impact
parameter of the scattering.  (iii) The quark-antiquark pair then become a $\rho$ long after they have passed the proton.  This is
the sequence of steps given in (51) where one also integrates over all possible dipole sizes and where the integration over impact
parameters gives a definite transverse momentum.

\begin{center}
\begin{figure}
\epsfbox[0 0 215 141]{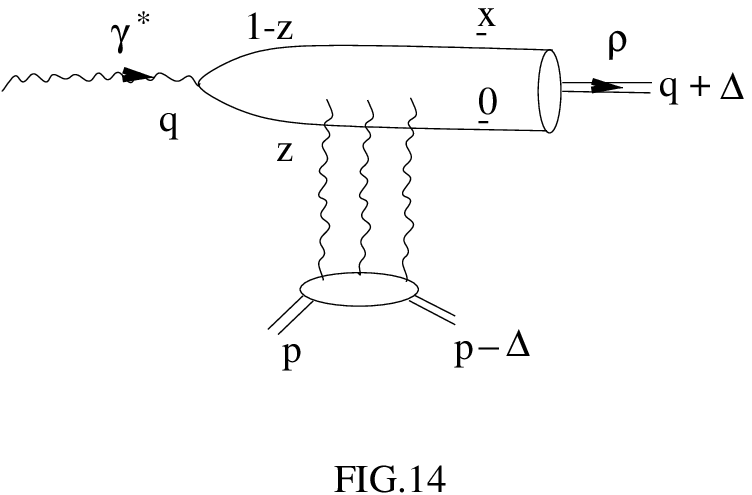}
\end{figure}
\end{center}

Now suppose $S$ is purely real, and define $N(Q) = (\psi_\rho,\psi_{\gamma^\ast}).$  Then one can take the square root of both ides of
(51), and after taking the inverse Fourier transform one finds

\begin{equation}
<S(x, r_0,b)> = 1 - {1\over 2N\pi^{3/2}}\int d^2\Delta e^{-i{\underline b}\cdot{\underline \Delta}}{\sqrt{{d\sigma\over dt}}}.
\end{equation}

\noindent In (52) $<S> = (\psi_\rho, S \psi_{\gamma^\ast})$ while $x$ denotes the Bjorken-x value of the scattering and $r_0(Q)$
denotes the typical value of the dipole size contributing to (51).  For (52) to have meaning it is important that the range of dipole
sizes contributing to (51) not be too large, and this appears to be the case for longitudinal $\rho$ production.  This analysis is
modeled on the classic analysis of Amaldi and Shubert\cite{Ama} for proton-proton elastic scattering.  Thus one can estimate the
$S-$matrix for a dipole of size $r_0,$ determined by $Q,$ scattering on a proton at a given impact parameter if the data are good
enough to carry out the integral in (52).  The analysis is not model independent as one needs to take a wavefunction for the
$\rho$\cite{Dos,Nem,fur}in order to evaluate $N.$  However, $N$ does not appear to be very sensitive to the choice of
wavefunction.  Also the data are not good enough at larger values of $t$ to accurately determine $S$ below $b \approx 0.3.fm.$  One
finds, for example, at $Q^2 = 3.5 GeV$ where $r_0 \approx 1/5 fm$ and for $x \approx 10^{-3}$ that: (i) $S(b \approx 0) \approx
0.5-0.7$; (ii) the probability of an inelastic collision $=1-S^2(b)$ is considerable at small values of $b$ indicating a reasonable 
amount of blackness at central impact parameters; (iii) $\sigma_{tot}^{q{\bar q}-Proton}\approx 10 mb$; (iv) ${\bar Q}_S$ is
consistent with that found in the Golec-Biernat W\"usthoff model.  This adds, perhaps, another piece of evidence that saturation is
approached in the HERA regime for moderate values of $Q^2.$

\section{A Simple Equation for Unitarity;\newline the Kovchegov equation$^{21}$}

Consider a (not too high momentum) dipole scattering on a high-energy hadron.  We suppose the quark-antiquark dipole is left moving
while the hadron is right moving.  Further we suppose that the rapidity, $y,$ of the dipole is such that $\alpha y << 1$ so that one
need not consider radiative corrections to the wavefunction of the dipole to evaluate the scattering amplitude.  We wish to study the
dependence of the elastic scattering amplitude as one changes the relative rapidity of the dipole and the hadron by an amount  $dY$
when the relative rapidity is $Y.$  Clearly one can view the change $dY$ either as increasing the momentum of the hadron and thus
allowing its wavefunction to evolve further or as increasing the momentum of the dipole.  The latter is easier to deal with since the
dipole is a simple object.  When the rapidity of the dipole is increased there is a small probability, proportional to $dY,$ that the
dipole emits a gluon before it scatters off the hadron.  We now calculate the probability for producing this quark-antiquark-gluon
state.

Since a gluon is emitted either off the quark or off the antiquark we have already done the basic emission amplitude, and it is given
in (8).  It will be convenient to work in a basis where transverse coordinate are used rather than transverse momenta so one must take
the Fourier transform of (8).  Thus the amplitude for a quark having transverse coordinate ${\underline x}_1$ to emit a gluon having
transverse coordinate ${\underline z},$ longitudinal momentum $k_-$ (Recall that in Sec.2 we were dealing with right movers while here
we are concerned with left movers.)  and polarization $\lambda$ is

\begin{equation}
\psi_\lambda^c({\underline z}-{\underline x}_1) = \int{d^2k\over(2\pi)^2}\ e^{i({\underline z}-{\underline x})\cdot{\underline
k}}\psi_\lambda^c(k)
\end{equation}

\noindent with $\psi_\lambda^c(k)$ given by (8) with the replacement $k_+\rightarrow k_-.$  We are now using $A_-=0$ light-cone gauge
to evaluate the left moving quark-antiquark-gluon state.  Using

\begin{equation}
\int {d^2k\over 2\pi} e^{i{\underline x}\cdot{\underline k}}{{\underline k}\over {\underline k}^2} = i {{\underline x}\over {\underline
x}^2}
\end{equation}

\noindent one gets

\begin{equation}
\psi_\lambda^c({\underline z}- {\underline x}_1)=({\lambda^c\over 2}) 2ig {({\underline z}-{\underline x}_1)\cdot {\underline
\epsilon}^{\lambda^\ast}\over ({\underline z}-{\underline x}_1)^2}{1\over{\sqrt{(2\pi)^32k_-}}}.
\end{equation}

\noindent To calculate the probability of a quark-antiquark-gluon state one adds the graphs in Fig.15 to get\cite{lle}

\begin{center}
\begin{figure}
\epsfbox[0 0 466 104]{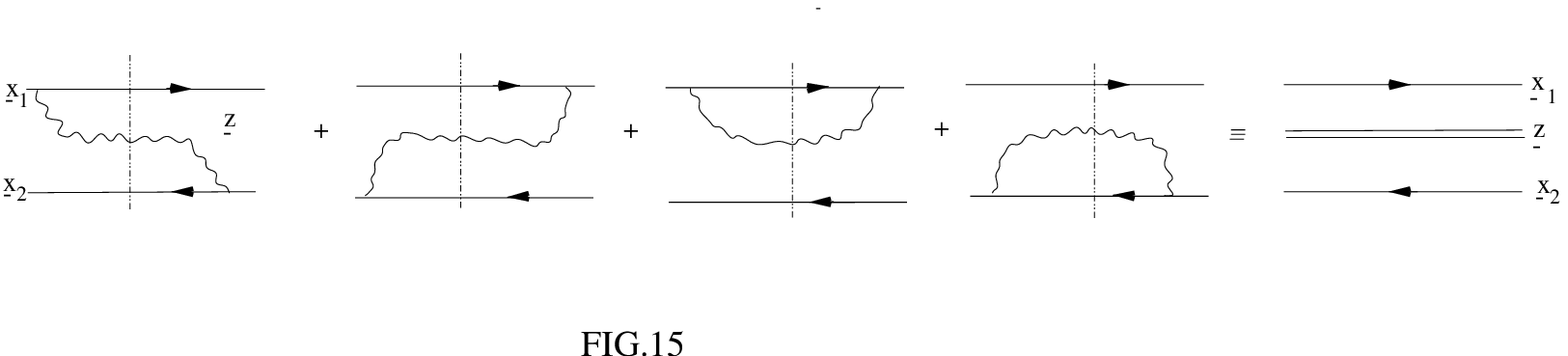}
\end{figure}
\end{center}

\begin{equation}
dP_r=\sum_c({\lambda^c\over 2}{\lambda^2\over 2})4g^2{d^2zdk_-\over (2\pi)^32k_-}[-2{({\underline z}-{\underline x}_1)\cdot({\underline
z}-{\underline x}_2)\over ({\underline z}-{\underline x}_1)^2({\underline z}-{\underline x}_2)^2} + {1\over ({\underline
z}-{\underline x}_1)^2} + {1\over ({\underline z}-{\underline x}_2)^2}]
\end{equation}

\noindent or

\begin{equation}
dP_r={\alpha N_c\over 2\pi^2} d^2z dY {({\underline x}_1-{\underline x}_2)^2\over ({\underline x}_1-{\underline z})^2({\underline
x}_2-{\underline z})^2}
\end{equation}

\noindent where we have set $dk_-/k_-=dY$ and $C_F={N_c\over 2}$ in the large $N_c$ limit where the Kovchegov equation will be valid.

Then the $S-$matrix for the quark-antiquark-gluon state to elastically scatter on the hadron multiplied by $dP_r$ in (57) gives the
change in the $S-$matrix, $dS,$ for dipole-hadron scattering.  The result is the Kovchegov equation\cite{gov}

\begin{displaymath}
{dS({\underline x}_1-{\underline x}_2,Y)\over dY} 
= {\alpha N_c\over 2\pi^2}\int d^2z {({\underline x}_1-{\underline x}_2)^2\over
({\underline x}_1-{\underline z})^2({\underline x}_2-{\underline z})^2}
\end{displaymath}
\begin{equation}
[S({\underline x}_1-{\underline z},Y) S({\underline z}-{\underline x}_2,Y)-S({\underline x}_1-{\underline x}_2,Y)],
\end{equation}

\noindent and it is illustrated in Fig.16.  We have assumed that the scattering of the two dipoles, the  quark-(antiquark part of the
gluon) and the (quark part of the gluon)-antiquark dipoles, factorize when scattering off the hadron.  This was clear in the model
Kovchegov considered where the hadron was a large nucleus.  This factorization is less obvious in the general case and the Kovchegov 
equation may be a sort of mean field approximation to a more complete equation.  Also the final term on the right-hand side of (58),
corresponding to the last two graphs in Fig.16, give the virtual contributions necessary to normalize the wavefunction\cite{lle}.  
The necessity of this last term can be seen by considering the weak interaction limit where $S\rightarrow 1.$  Then the final term on
the righthand side of (58) is necessary to get ${dS\over dY}=0$ when $S=1.$

\begin{center}
\begin{figure}
\epsfbox[0 0 337 113]{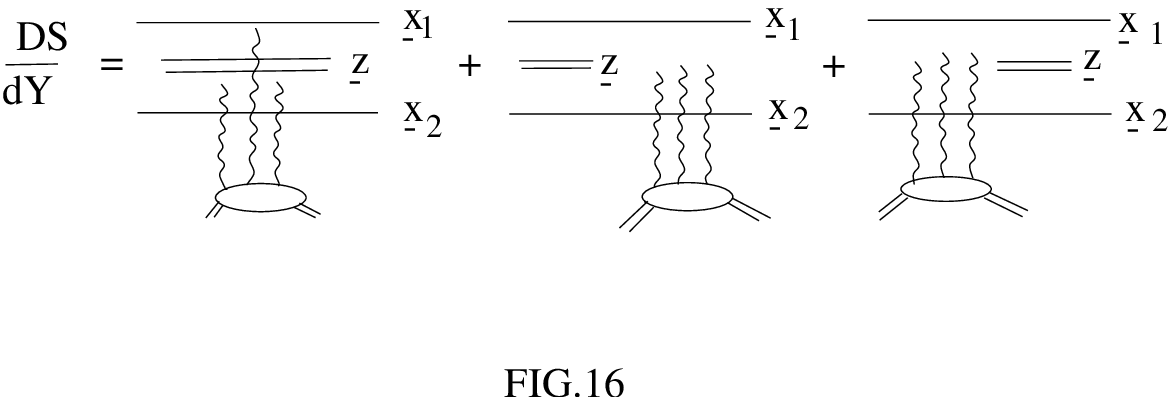}
\end{figure}
\end{center}

There are two interesting limits to (58).  First suppose that $S$ is near 1 and write $S=1-iT.$  One easily finds, keeping only
linear terms in T,

\begin{equation}
{dT({\underline x}_1-{\underline x}_2,Y)\over dy} = {\alpha N_c\over \pi^2} \int d^2z {({\underline x}_1-{\underline x}_2)^2\over
({\underline x}_1-{\underline z})^2({\underline x}_2-{\underline z})^2}[T({\underline x}_1-{\underline z},Y) - {1\over 2} T({\underline
x}_1-{\underline x}_2, Y)]
\end{equation}

\noindent which is the dipole form of the BFKL equation.  In this case the factorized form of the scattering is justified by the
large $N_c$ limit and the weak coupling approximation.

The other interesting limit is where $S$ is small in which case one need only keep the second term on the righthand side of (58) giving

\begin{equation}
{dS({\underline x}_1-{\underline x}_2,Y)\over dY} = - {\alpha N_c\over 2\pi^2}\int {d^2z({\underline x}_1-{\underline x}_2)^2\over
({\underline x}_1-{\underline z})^2({\underline x}_2-{\underline z})^2} S({\underline x}_1- {\underline x}_2Y).    
\end{equation} 

\noindent Of course (60) as written cannot be valid.  The assumption that  $S$  be small can be true only when the dipole size is large
compared to $1/Q_s.$  Thus we should restrict the integration in (60) to the region $({\underline x}_1-{\underline x}_2)^2 >>
1/Q_S^2$   as well as to the region $({\underline x}_1-{\underline z})^2, ({\underline x}_2-{\underline z})^2 >> 1/Q_S^2$ so that the
nonlinear term in (58) not cancel the linear term.  In the logarithmic regions of integration one can rewrite the integral (60) as

\begin{equation}
{dS({\underline x}_1-{\underline x}_2,Y)\over dY} = - 2 {\alpha N_c\over 2\pi^2} \int_{1/Q_S^2}^{({\underline x}_1-{\underline
x}_2)^2} {\pi d({\underline x}_1-{\underline z})^2\over ({\underline x}_1-{\underline z})^2}S({\underline x}_1-{\underline x}_2,Y)
\end{equation}

\noindent giving

\begin{equation}
{dS({\underline x},Y)\over dY} = - {\alpha N_c\over \pi} \ell n (Q_S^2{\underline x}^2) S({\underline x},Y)
\end{equation}

\noindent whose solution is

\begin{equation}
S({\underline x}, Y) = e^{-{\alpha N_c\over \pi}\int_{Y_0}^Y dy\   \ell n[Q_S^2(y){\underline x}^2]}S({\underline x},Y_0).
\end{equation}

\noindent If $Q_S^2$ is exponentially behaved

\begin{displaymath}
Q_S^2(y) = e^{c{\alpha N_c\over \pi}(y-Y_0)}Q_S^2(Y_0)
\end{displaymath}

\noindent then

\begin{equation}
S({\underline x},Y) = e^{-{c\over 2}({\alpha\cdot N_c\over \pi})^2(Y-Y_0)^2}S({\underline x},Y_0)
\end{equation}

\noindent where $Y_0$ should be chosen to satisfy

\begin{equation}
Q_S^2(Y_0) {\underline x}^2 = 1.
\end{equation}

\noindent Eq.(64) is the result found in Ref.23 and argued heuristically already some time ago\cite{Sal}, although without an
evaluation of the coefficient of the $Y^2$ term in the exponent.

The Kovchegov equation is an interesting equation for studying scattering when one is near unitarity limits.  In the next section we
shall use a procedure very similar to what we have done here to derive an equation whose content is presumably equivalent to the
Balitsky equation\cite{Bal}, a somewhat more general form than the Kovchegov equation.  The advantage of (58) is its simplicity and
its likely qualitative correctness in QCD.

\section{A Simple Derivation of the\newline JIMWLK equation$^{[14,25-27]}$}

Over the past seven years or so there has been an ambitious program dedicated to finding appropriate equations for dealing with high
density wavefunctions in QCD.  This program has been quite successful and a  renormalization group equation in the form of a
functional Fokker-Planck equation for the wavefunction of a high-energy hadron has been given by the authors of Refs.14 and 25-27. 
(JIMWLK).  The most complete derivation is given in Ref.27 where the equation is written in terms of a covariant gauge potential,
$\alpha,$ coming from light-cone gauge quanta in a high-energy hadron.  Here we give an alternative simple derivation\cite{uel}.

We can imitate this mixture of gauges used in Ref.27 by taking Coulomb gauge\cite{ros} which has a gluon propagator

\begin{equation}
{\large D}_{\alpha\beta}(k) = - {i\over k^2}[g_{\alpha\beta}-{N\cdot k(N_\alpha k_\beta + N_\beta k_\alpha)-k_\alpha
k_\beta\over ({\vec k})^2}]
\end{equation}

\noindent where $N\cdot v = v_0$ for any vector $v.$  Suppose the propagator has $k_+^2 >> {\underline k}^2, k_-^2$ and connects two
highly right moving lines.  Then the Coulomb gauge propagator is equivalent to the $A_+=0$ light-cone gauge propagator

\begin{equation}
{\large D}_{\alpha\beta}(k) = {-i\over k^2}[g_{\alpha\beta}-{\eta_\alpha k_\beta + \eta_\beta k_\alpha\over \eta\cdot k}]
\end{equation}

\vskip 5pt
\noindent{\bf Problem 12(E)}:  Show that ${\large D}_{--},$ and ${\large D}_{-i}$ as given in (66) and (67) agree when
$ {\underline k}_+^2 >> {\underline k}^2, k_-^2.$  For a right moving
system these are the important components of the gluon propagator.

Similarly for a left moving system Coulomb gauge is equivalent to $A_-=0$ gauge while
for gluon lines which connect left moving systems to right moving
systems the dominant component in (66) is $D_{+ -} = 1/({\vec k})    ^2$ which
looks like covariant gauge when $k_+^2, k_-^2 << {\underline k}^2.$  

We are going to consider the scattering of a set of left moving quanta,
quarks and gluons, on some high-energy, right moving hadron.  These quarks
and gluons may be parts of a hadronic wavefunction which are frozen in the
passage over the right moving hadron or they may come from a current as in
our discussion of deep inelastic scattering given in Sec.8.  For
simplicity we shall limit our discussion to left moving quark and
antiquark lines, but this is simply to avoid too cumbersome notation. 
Then a left moving quark interacting with the right moving hadron can be
represented by

\begin{equation}
V^\dagger({\underline x}) = P\  exp\{ig\int_{-\infty}^\infty dx_-A_+({\underline x}, x_-)
\end{equation}

\noindent where we have taken $x_+=0$ and fixed the left moving quark to have transverse coordinate ${\underline x}.$  Except for the
change of right moving quark to left moving quark (68) is the same as the operator in the matrix element in (26) where we showed how
 quarks could be identified with Wilson lines in the fundamental representation.  By taking gauge invariant combinations of
$V$'s and $V^\dagger$'s we can form observables which depend on $A_+$ and which correspond to the scattering of quite general
left moving systems on the right moving hadron.  We denote a general such observable by $O(A_+).$

Although we have put the integration in (68) exactly on the light-cone we in fact are going to assume that the left moving observable
has rapidity $y$ obeying $\alpha y << 1$ so that transverse gluons are unlikely to be emitted by the left movers allowing us to
identify the left moving system at $x_-=\pm \infty.$  If the relative rapidity of the scattering is $Y$ then we imagine that
$\alpha(Y-y) \approx \alpha Y >> 1$ so that the right moving hadron has, in general, a wavefunction including many gluons.  If the
right moving hadron has momentum  $p$  then the scattering amplitude is

\begin{equation}
<O>_Y = <p\vert O\vert p > = \int {\large D}[\alpha({\underline x}, x_-)]O(\alpha) W_Y[\alpha]
\end{equation}

\noindent where the weight function $W_Y$ is given by

\begin{equation}
W_Y[\alpha]=\int D[A_\mu]\delta(A_+-\alpha)\delta(F(A))\Delta_F[A] e^{iS[A]}
\end{equation}

\noindent where  $D$ indicates a functional integral, $F$ is a gauge fixing, and $\Delta_F$ is the corresponding
Fadeev-Popov determinant times an operator which projects out the state $\vert p >$ initially and finally.  We suppose $W_Y$ is
normalized to

\begin{equation}
\int  D[\alpha]W_Y[\alpha]=1.
\end{equation}

Now consider the $Y-$dependence of $< O >_Y.$  From (69) one can clearly write

\begin{equation}
{d\over dY}  <O>_Y = \int D[\alpha]O(\alpha){d\over dY} W_Y[\alpha].
\end{equation}

\noindent However, one can equally well imagine calculating ${d\over dY}<O>_Y$ by evaluating the change of the left moving system, due
to an additional gluon emission, as one increases the rapidity of the left moving system by an amount $dY.$  The change in the
left moving state is given by the gluon emissions and absorptions shown in Figs.17 and 18 where spectator quark and antiquark lines are
not shown. In Fig.17 we have assumed that a gluon connects a quark and an antiquark line.  This is for definiteness.   We equally
well  could have assumed a connection to two quark lines or to two antiquark lines.  The vertical line in the figures represents the
``time,'' $x_-=0,$ at which the left moving system passes the right moving hadron. This view of the $Y-$dependence of $<O>_Y$ in terms
of a change of rapidity of the (rather simple) left movers is in the spirit of work previously done by Balitsky[5],
Kovchegov\cite{gov}, and Weigert\cite{Wei}.

\begin{center}
\begin{figure}
\epsfbox[0 0 456 127]{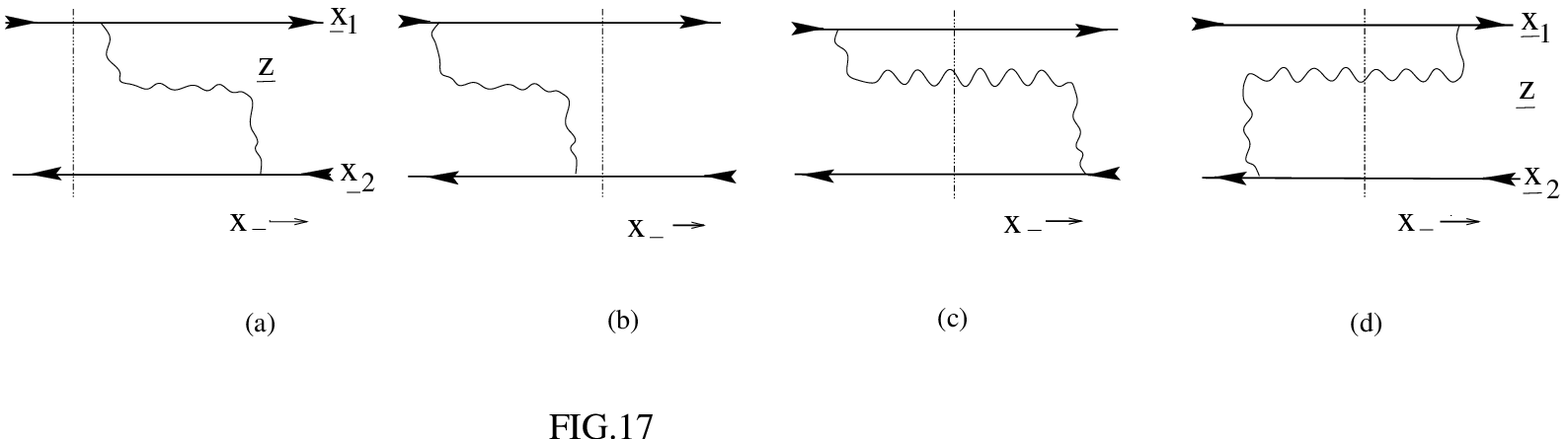}
\end{figure}
\end{center}

We shall examine in some detail the graphs of Fig.17 before stating the complete result including the graphs of Fig.18.  Begin with
the graph shown in Fig.17c. We do the  calculation in $A_-=0$ light-cone gauge, which for left movers is equivalent to our Coulomb
gauge choice.  This graph is exactly the same as has been calculated in Sec.11 except for the fact that the quark and antiquark lines
are not necessarily in a color singlet.  The result is the same as ${1\over 2}$ the first term on the righthand side of (56) except
for the color factors which we put in separately.  The result is

\begin{center}
\begin{figure}
\epsfbox[0 0 936 94]{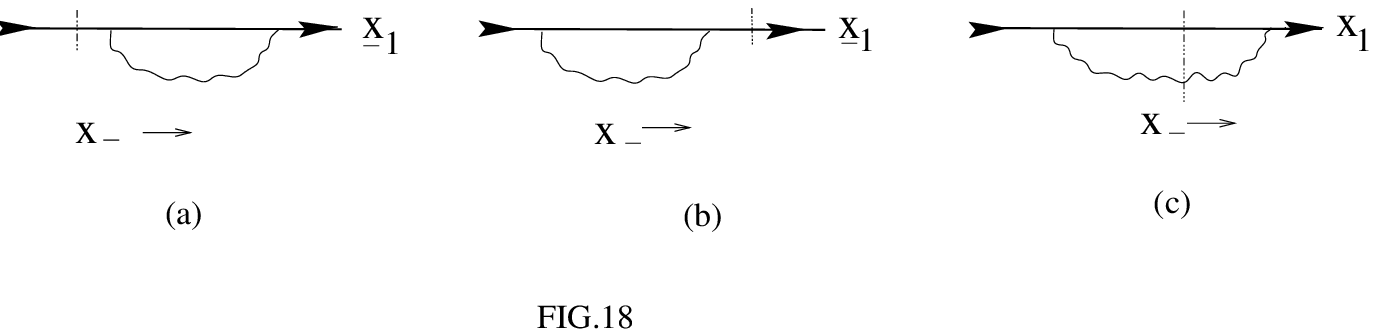}
\end{figure}
\end{center}

\begin{equation}
V^\dagger({\underline x}_1)\otimes\ V({\underline x}_2)\to -{\alpha_S\over \pi^2}\int d ^2z{({\underline x}_1-{\underline
z})\cdot({\underline x}_2-{\underline z})\over ({\underline x}_1-{\underline z})^2({\underline x}_2-{\underline
z})^2} {\tilde V}_{cd}({\underline z})V^\dagger({\underline x}_1)T^c\otimes\ V({\underline x}_2)T^d
\end{equation}

\noindent in going from $O$ to ${dO\over dY}.$  The additional factors, as compared to the corresponding term in (56), are $T^c$
which comes to the right of $V^\dagger({\underline x}_1)$ because the emission off the ${\underline x}_1-$ line is  at early values of
$x_-,$ the $T^d$ which comes to the right of $V({\underline x}_2)$ because the absorption on the ${\underline x}_2-$line is at late
values of $x_-,$ and the factor ${\tilde V}_{cd}({\underline z})$ giving the interaction of the gluon with the hadron as a Wilson
line in the adjoint representation.

Now write

\begin{equation}
V^\dagger({\underline x}_1)T^c=V^\dagger({\underline x}_1)T^cV({\underline x}_1)V^\dagger({\underline x}_1)={\tilde
V}_{ca}({\underline x}_1)T^aV^\dagger({\underline x}_1).
\end{equation} 

\noindent Now $T^a$ comes to the left of $V^\dagger({\underline x}_1)$ as if the emission of the gluon were at late values of $x_-.$ 
Indeed, we may view the graph in Fig.17c as being given by the ``mnemonic'' graph shown in Fig.19 where the adjoint line integral
starts at large positive values of $x_-$ and proceeds to large negative values of $x_-$ at a transverse coordinate ${\underline x}_1$
then back again to large positive values of $x_-$ at a transverse coordinate  ${\underline z}.$  The result of the graph of Fig.17c
then is

\begin{center}
\begin{figure}
\epsfbox[0 0 201 145]{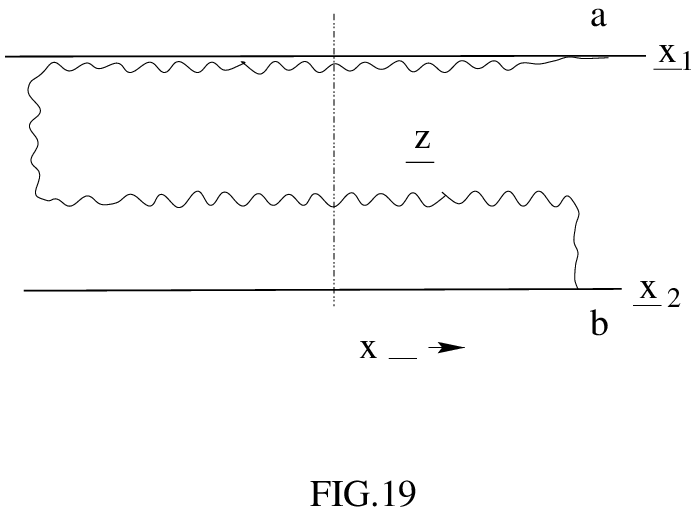}
\end{figure}
\end{center}

\begin{displaymath}
V^\dagger({\underline x}_1)\otimes\ V({\underline x}_2)\rightarrow {\alpha_S\over \pi^2}\int d^2z{({\underline x}_1-{\underline
z})\cdot({\underline x}_2-{\underline z})\over ({\underline x}_1-{\underline z})^2({\underline x}_2-{\underline
z})^2}\{-{\tilde V}^\dagger(x_1){\tilde V}({\underline z})\}_{ab}
\end{displaymath}
\begin{equation}
T^aV^\dagger({\underline x}_1)\otimes V({\underline x}_2)T^b
\end{equation}

Now it is straightforward to add in the other terms of Fig.17 and those of Fig.18 to get

\begin{displaymath}
V^\dagger({\underline x}_1) \otimes V({\underline x}_2) \rightarrow \alpha_S\{{1\over 2}\int d^2xd^2y \eta_{{\underline
x}{\underline y}}^{ab}{\delta^2\over \delta\alpha^a({\underline x},x_-)\delta\alpha^b({\underline y},y_-)} +\int d^2x \nu_{\underline
x}^a{\delta\over \delta\alpha^a({\underline x},x_-)}\}\cdot
\end{displaymath}

\begin{equation}
\cdot V^\dagger({\underline x}_1)\otimes\ V({\underline x}_2)
\end{equation}

\noindent where, if one takes large positive values of $x_-$ and $y_-,$ the functional derivatives in (76) simply insert color
matrices in the appropriate places, as, for example, in (75).  $\eta$ and $\nu$ are given by

\begin{displaymath}
g^2\eta_{{\underline x}{\underline y}}^{ab}=4\int {d^2z\over 4\pi^2}\ {({\underline x}-{\underline z})\cdot({\underline y}-{\underline
z})\over ({\underline x}-{\underline z})^2({\underline y}-{\underline z})^2}\cdot
\end{displaymath}
\begin{equation}
\cdot \{1+{\tilde V}^\dagger({\underline x}){\tilde V}({\underline y})-{\tilde V}^\dagger({\underline x}){\tilde V}({\underline
z})-{\tilde V}^\dagger({\underline z }){\tilde V}({\underline y})\}_{ab}
\end{equation}

\noindent and

\begin{equation}
g\nu_{\underline x}^a=2i\int {d^2z\over ({\underline x}-{\underline z})^2} tr [T^a{\tilde V}^\dagger({\underline
x}_1){\tilde V}({\underline z})].
\end{equation}

For a general scattering one simply replaces $V^\dagger({\underline x}_1)\otimes V({\underline x}_2)$ in (76) by $O(\alpha).$  After
multiplying by $W_Y[\alpha]$ and integrating over ${\em D}[\alpha]$ one gets\cite{uel}

\begin{displaymath}
\int D[\alpha]O(\alpha){dW_Y[\alpha]\over dY} = \int D[\alpha]O(\alpha)\alpha_S\{{1\over 2} d^2xd^2y{\delta^2\over
\delta\alpha^a({\underline x},x_-)\delta\alpha^b({\underline y},y_-)}
\end{displaymath}

\begin{equation}
\cdot[W_Y\eta_{{\underline x}{\underline y}}^{ab}]-\int d^2x {\delta\over \delta\alpha^a({\underline x},x_-)}[W_Y\nu_{\underline
x}^a]\}
\end{equation}

\noindent where an integration by parts in $\alpha$ has been done on the  righthand side of (79).  To the extent that the $O(\alpha)$
form a complete set of observables, and it is not clear how close this is to being true, one can equate the  integrands of (79) and
arrive at the JIMWLK equation

\begin{displaymath}
{dW_Y[\alpha]\over dY} = \alpha_S\{{1\over 2}\int d^2x d^2y {\delta^2\over \delta \alpha^a({\underline
x},x_-)\delta \alpha^b({\underline y},y_-)}
[W_Y\eta_{{\underline x}{\underline y}}^{ab}]
\end{displaymath}
\begin{equation}
 -\int d^2x {\delta\over \delta\alpha^a({\underline x}, x_-)}[W_Y\nu_{\underline x}^a]\}.
\end{equation}

The exact values of $x_-$ and $y_-$ appear to have some arbitrariness as discussed in some detail in Ref.28.  Eq.(80) is an elegant
equation of a functional Fokker-Planck type the nature of whose solutions is now under investigation.


\begin{thebibliography}{99}
\bibitem{Yu}	Yu.V. Kovchegov,{\em Phys.Rev.D}{\bf 54} (1996) 5463; {\bf D55} (1997) 5445.
\bibitem{Kov} 	Yu. V. Kovchegov and A.H. Mueller,{\em Nucl.} Phys.{\bf B529} (1998) 451.
\bibitem{Buc} 		W. Buchm\"uller, M.F. McDermott and A. Hebecker, {\em Nucl.Phys.} {\bf B487} (1997) 283; {\bf B500} (1997) 621 (E)
\bibitem{Geh}	 	W. Buchm\"uller, T. Gehrman and A. Hebecker, {\em Nucl.Phys.} {\bf B538} (1999) 477.
\bibitem{Bal}	 	I. Balitsky, {\em Nucl. Phys.}{\bf B463} (1996) 99.
\bibitem{Mue}			A.H. Mueller, {\em Nucl.Phys.} {\bf B558} (1999) 285.
\bibitem{Bjo} 		J.D. Bjorken in {\em Proceedings of the International Symposium on Electron and Photon Interactions at High
Energies}, pages 281-297, Cornell (1971).
\bibitem{ler}		 A.H. Mueller, {\em Nucl. Phys.}{\bf B335} (1990) 115.
\bibitem{Fra}		 L.L. Frankfurt and M. Strikman, {\em Phys.Rep.}{\bf 160} (1998) 235.
\bibitem{tel}			B. Bl\"attel, G. Baym, L.L. Frankfurt and M. Strikman, {\em Phys. Rev.Lett.} {\bf 70} (1993) 896.
\bibitem{Zak}			B.G. Zakharov, {\em JETP Lett.}{\bf 63} (1996) 952.
\bibitem{ran}		L.McLerran and R. Venugopalan,{\em Phys.Rev.}{\bf D49}(1994) 2233;{\bf D49}(1994) 3352; {\bf D50}(1994) 2225.
\bibitem{Jal}	J. Jalilian-Marian, A. Kovner, L. McLerran and H. Weigert, {\em Phys. Rev.} {\bf D55} (1997) 5414.
\bibitem{Ian}		E. Iancu and L. McLerran,Phys.Lett.{\bf B510} (2001) 145.
\bibitem{Gol}		K. Golec-Biernat and M. W\"usthoff,{\em Phys. Rev.}{\bf D59} (1999) 014017; {\em Phys.Rev.}{\bf D60} (1999) 114023.
\bibitem{Mun}		S. Munier, A.M. Stasto and A.H. Mueller, Nucl.Phys. {\bf B603} (2001) 427.
\bibitem{Ama}		U. Amaldi and K. R. Schubert,{\em Nucl.Phys.}{\bf B166} (1980) 301.
\bibitem{Dos}		H.G. Dosch, T. Gousset, G. Kulzinger and H.J. Pirner, {\em Phys. Rev.}D{\bf 55} (1997) 2602.
\bibitem{Nem}	J. Nemchik, N.N. Nikolaev, E. Predazzi and B.G. Zakharov,{\em Z.Phys.}{\bf C75} (1997) 71.
\bibitem{fur}	L. Frankfurt, N. Koepf and M. Strikman, {\em Phys.Rev.}D{\bf 54} (1996) 3194.
\bibitem{gov}	Yu. Kovchegov, {\em Phys.Rev.}{\bf D60} (1999) 034008; {\em Phys.Rev.}{\bf D6}1 (2000) 074018.
\bibitem{lle}	A.H.Mueller, {\em Nucl. Phys.} {\bf B415} (1994) 373.
\bibitem{Lev}	E. Levin and K. Tuchin, {\em Nucl.Phys.}{\bf B573} (2000) 83; hep-ph/0101275.
\bibitem{Sal}	A.H. Mueller and G.P. Salam, {\em Nucl.Phys.}{\bf B475} (1996) 293.
\bibitem{Mar}	J. Jalilian-Marian, A. Kovner, A. Leonidov and H. Weigert, {\em Nucl. Phys.}{\bf B504} (1997) 415; Phys. Rev. {\bf D59}
(1999) 014014.
\bibitem{Wei}	H. Weigert, hep-ph/0004044.
\bibitem{Leo}		E. Iancu, A. Leonidov and L. McLerran, {\em Nucl.Phys.} {\bf A692} (2001) 583; {\em Phys.Lett.}{\bf B510} (2001)
133; hep-ph/0109115.
\bibitem{uel} A.H. Mueller, hep-ph/0110154.
\bibitem{ros}	T. Jaroszewicz, {\em Acta Physica Polonica} {\bf B11} (1980) 965.
\end{thebibliography}
\end{document}